\documentclass[twocolumn,apj]{emulateapj}

\usepackage{natbib}
\usepackage{graphicx}
\usepackage{subfigure}

\usepackage{amsmath,amssymb}
\usepackage{ulem}

\newcommand{\be}{\begin{equation}}
\newcommand{\ee}{\end{equation}}

\begin{document}

\title{Three-dimensional Iroshnikov-Kraichnan turbulence in a mean magnetic field
}
\author{Roland~Grappin}
\email[]{Roland.Grappin@lpp.polytechnique.fr}
\affiliation{LUTH, Observatoire de Paris and LPP, Ecole Polytechnique
}

\author{Wolf-Christian~M\"{u}ller}
\email[]{Wolf-Christian.Mueller@tu-berlin.de}
\affiliation{Technische Hochschule Berlin, Zentrum f\"ur Astronomie und Astrophysik, Germany}


\author{Andrea Verdini}
\affiliation{Universit\`a di Firenze, Dipartimento di Fisica e Astronomia, Firenze, Italy
and Royal Observatory of Belgium, SIDC/STCE, Brussels}

\author{\"Ozg\"ur G\"urcan}
\affiliation{LPP, Ecole Polytechnique}

\date{\today}

\begin{abstract}
Forced, weak MHD turbulence with guide field is shown to adopt different regimes, depending on the magnetic excess of the large forced scales. When the magnetic excess is large enough, the classical perpendicular cascade with $5/3$ scaling is obtained, while when equipartition is imposed, an isotropic $3/2$ scaling appears in all directions with respect to the mean field (\cite{2010PhRvE..82b6406G} or GM10).
We show here that the $3/2$ scaling of the GM10 regime is not ruled by a small-scale cross-helicity cascade, and propose that it is a 3D extension of a perpendicular weak Iroshnikov-Kraichnan (IK) cascade.
We analyze in detail the structure functions in real space and show that they closely follow the critical balance relation both in the local frame and the global frame: we show that there is no contradiction between this and the isotropic $3/2$ scaling of the spectra.
We propose a scenario explaining the spectral structure of the GM10 regime, that starts with a perpendicular weak IK cascade and extends to 3D by using quasi-resonant couplings. The quasi-resonance condition happens to reduce the energy flux in the same way as is done in the weak perpendicular cascade, so leading to a $3/2$ scaling in all directions.
We discuss the possible applications of these findings to solar wind turbulence.
\end{abstract}

\maketitle

\section{Introduction}
Plasma turbulence in a mean magnetic field corresponds to a rather ubiquitous astrophysical setting. 
Understanding its nonlinear dynamics which gives rise to measurable two-point statistics such as the energy spectra and structure functions is thus highly important. 
The current state of affairs with regard to the simplified incompressible magnetohydrodynamic approximation may be summarized as follows.
When increasing the mean field $B_0$, turbulence becomes mainly 2-dimensional, namely, gradients develop mainly in the direction perpendicular to the mean field.

There are presently two known classes of turbulent regimes: 
(i) the weak IK regime \citep{Iroshnikov:1963p9274,Kraichnan:1965p9279} with a $k^{-3/2}$ energy spectrum due to 
repeated random uncorrelated collisions of Alfv\'en wavepackets - this regime shows a $k^{-3/2}$ scaling for the energy spectrum; it has been found in purely 2D configurations with no mean field within the plane \citep{Pouquet:1988dv,Biskamp:1989ug}; (ii) truly 3D regimes that tend to become quasi 2D as the mean field increases.
There is no possibility to go smoothly from the 3D/quasi 2D regimes to the 2D IK regime which appears to be an isolated island.

The tendency of 3D regimes to become quasi-2D is a direct consequence of
(i) the nonlinear terms that couple oppositely propagating Alfv\'en species; (ii) the dominance of resonant interactions that select couplings driving a cascade perpendicular to the mean field \citep{Montgomery:1981p6103,Shebalin:1983p6056,Grappin:1986p707}.

A numerical realization of the progressive two-dimensionalization of turbulence is provided by the work of \cite{Muller:2003p809}. 
In these experiments, large-scale flows and magnetic fields with equal energies are imposed, within the spectral range $1Ê\le k \le 2$, isotropically in all directions, and the aspect ratio of the numerical domain (with periodic boundaries) is unity.
Two-dimensionalization is revealed by measuring Fourier space anisotropy: energy in the parallel direction is seen to decrease as $B_0$ increases (\cite{Muller:2003p809}, \cite{2010PhRvE..82b6406G}.

Another property found by \cite{Muller:2003p809} is that the spectral index in the dominant (perpendicular) direction becomes increasingly close to $-3/2$ when $B_0$ is increased from $5 b_{rms}$ to $10 b_{rms}$.
The aspect ratio is found to increase when scale decreases (by measuring the slope of structure functions).
More precisely, the parallel and perpendicular real space scalings (as found by structure functions) follow closely the prediction of critical balance between nonlinear time and linear parallel Alfv\'en propagation time
(\cite{Goldreich:1995p4882} (GS)), with the caveat that the dominant (perpendicular) scaling is somewhat flatter, corresponding to a spectral index of $-3/2$ instead of $-5/3$. In Fourier space on the contrary, all directions show the same $-3/2$ scaling, 
with all the anisotropy reducing to an amplitude anisotropy of large scale fluctuations (GM10): this implies a high level of excitation at parallel/oblique small scales, at odds with predictions of critical balance.

Solar wind turbulence shows similar properties, in common with the GM10 regime just described: observations report anisotropic SF scaling laws that are compatible with the GS predictions \citep{2008PhRvL.101q5005H}; at the same time, the observed cosmic ray diffusion requires the existence of a substantial excitation of the (not directly measurable) 3D power-spectrum along the parallel direction near $k_\bot=0$ \citep{2000PhRvL..85.4656C}, which is not a direct consequence of the critical balance phenomenology.

In this paper, we show how the properties of the largest scales control  
the extension of the spectral energy distribution along the parallel direction, as well as the spectral scaling.
We show that a large parallel/oblique extent of the spectrum, together with the $-3/2$ scaling characteristic of the GM10 regime is attained when magnetic/kinetic energy ratio is close to unity at large scales, while the $-5/3$ scaling together with a purely perpendicular cascade results when the large scale magnetic/kinetic energy ratio is large.

We furthermore provide a scenario to explain how the system passes from one state to the other, namely from a preexisting perpendicular $k_\bot^{-3/2}$ cascade to an oblique one. Only in the case where the perpendicular slope is $-3/2$ do we obtain an isotropic scaling.
The scenario is able to reproduce as well the amplitude anisotropy between the parallel and perpendicular directions. We compare in the discussion the present situation with non zero mean field to the one existing in the zero mean field case, where the scaling properties seem as well to be controlled by large-scale properties. 

Our new turbulence scenario fills a missing link. Previously, the IK turbulence had been obtained in 2D simulations (with no mean field), as an isolated, singular regime. The 3D MHD turbulence was, asymptotically at least (i.e., at small scales) the domain of strong turbulence. 
No link was proposed between the two regimes. On the contrary, the new 3D turbulent regime we propose tends to the IK 2D turbulence when the global mean field goes to infinity; it is an anisotropic 3D extension of the IK regime.

The plan of this article is as follows. In section 2, we recall the existing phenomenologies. In section 3, we reanalyze the properties of the numerical simulations studied in \cite{2010PhRvE..82b6406G}. In section 4, we describe the ricochet process generating the oblique cascade.
The last section is a discussion.

\section{Equations and phenomenologies}

The MHD equations read, written in terms of
the so-called Elsasser variables $z^\pm = u\mp b$:
\be
\partial_t \textbf z^\pm \pm (\textbf B_0 \cdot \nabla) \textbf z^\pm + (\textbf z^\mp \cdot \nabla) \textbf z^\pm + \nabla P = 0
\label{base1}
\ee
with $\nabla \cdot \textbf z^\pm=0$, and where we leave the diffusive terms aside. 
After taking the Fourier transform, this leads to
\be
\partial_t \widehat{z^\pm_i}(k) \pm ik_\parallel B_0 \widehat{z^\pm_i}(k) + i P_{ijl}(k)   \widehat{z^\pm_l z^\mp_j}(k)
= 0
\label{MHD0}
\ee
where $k_\parallel$ denotes the projection of the wave vector on the mean field $B_0$ and $P_{ijl}(k)$ is the usual projection operator $P_{ijl}(k) = k_j(\delta_{il} - k_lk_i/k^2)$.
We go on and rewrite the equations using the Heisenberg representation for the unknown amplitudes:
$\widehat{\textbf z^\pm}(k) = \widehat{\textbf u^\pm}(k)e^{\pm ik_\parallel B_0 t}$.
After rearranging the different factors $e^{\pm i(k_\|,p_\|,q\|)  B_0 t}$, taking into account the condition $k=p+q$ we obtain
\be
\partial_t \widehat{u_i^\pm}(k) + iP_{ijl}(k)\int  d^3q \ \widehat{u_l^\pm}(p)\widehat{u_j^\mp}(q) e^{\mp 2i q_\| B_0 t} = 0
\label{Upm}
\ee
where in the integrand $p$ stands for $k-q$, i.e., the triadic relation $k=p+q$ is always respected.
(See  \cite{Grappin:1986p707} for a discussion of the equation in the 2D case).
This equation for the wave amplitudes $\widehat{u^\pm}$ has the merit of 
separating clearly the nonlinear coupling from the linear coupling due to the mean field $B_0$: 
while the former redistributes amplitudes all along the spectrum (i.e., coupling two wave vectors $p$ and $q$,
so changing the amplitude of the wave vector $k$), the latter decreases strongly the effect of the former,
as soon as the oscillating term is dominant.

In the following, we will use the following simplified version of this equation, replacing the complicated kernel $P_{ijl}$ by a dimensional factor $k$ and not distinguishing among field components. The simplified version of eq.~\ref{Upm} for
$\widehat{u^\pm_k}$ reads:
\begin{equation}
\partial_t \widehat{u^\pm_k}  = k \int d^3q \ \widehat{u^\pm_p} \widehat{u^\mp_q} e^{\mp 2i q_\| B_0 t}
\label{base2}
\end{equation}

\subsection{The Kolmogorov cascade}
We give here a short derivation of the Kolmogorov scaling in the hydrodynamic case of the Euler equation ($u^+ = u^- = u$, $B_0=0$), which will serve as a guide in the following. 
Denote by $\ell$ a scale between the largest scale and dissipation scale, and $u_\ell$ the
typical value of the velocity associated to scales $\sim l$. 
A correct definition is the r.m.s value of the velocity subject to bandpass filtering, say of an octave around the wavenumber $k=1/\ell$ \citep{1995tlan.book.....F}. A working definition \citep{Rose:1978wv} is to take 
\be
u_\ell^2/2 \simeq \int_{k/\sqrt 2}^{k\sqrt 2} E(k') dk' \simeq k E(k).
\label{rose}
\ee
where $E(k)=4\pi k^2 |\widehat{u_k}|^2$ is the 1D spectral energy density.
The eddy turnover (or non-linear) time is 
\be
t_\ell = \ell/u_\ell.
\label{tnl}
\ee
It is the typical time taken for a structure of size $\sim \ell$ to undergo a significant distortion due to the relative motion $u_\ell$, and thus, as well, the time for energy to be transferred from scales $\sim \ell$ to smaller scales. 
The energy flux from scale $\ell$ to smaller scales thus may be estimated as
\be
F_\ell \sim u^2_\ell/t_\ell \sim u^3_\ell/\ell 
\label{fk41}
\ee
In the inertial range, dissipation may be neglected, and since the kinetic energy is an inviscid invariant, 
 the energy flux does not depend on scale $\ell$.
The constraint $F_\ell = const = \epsilon$, leads to the velocity $u_\ell$ scaling as
\be
u_\ell \sim \epsilon^{1/3} \ell^{1/3}
\label{k41}
\ee
which implies, in view of eq.~\ref{rose}, that the energy spectrum scales as
\be
E(k) \sim \epsilon^{2/3} k^{-5/3}
\ee
In the following, we will use systematically the wavenumber $k$ as an index, instead of the scale $\ell=1/k$.

The main hypothesis at the basis of the Kolmogorov phenomenology 
is the localness hypothesis, which means that the convolution integral in eq.~\ref{Upm}-\ref{base2} mainly couples wavevectors close one to the other:
\be
|k| \sim |p| \sim |q|
\label{localness}
\ee
where the sign $\sim$ must be interpreted as ``equal within a factor two''.
This localness hypothesis is required to express the convolution integral in terms of the sole wavenumber $k$, or as well, the single scale $\ell$, leading to the simple form in eq.~\ref{fk41}
for the energy flux.
It will be used repeatedly in the following.
The localness hypothesis actually allows to derive the characteristic time scale in a straightforward way from the primitive equation. The primitive eq.~\ref{base2} may be rewritten formally when $B_0=0$ and $u^+ = u^-$ as
\be
\partial_t u_k = k u_k u_k
\label{ukuk}
\ee
which leads again to the eq.~\ref{tnl} for the nonlinear time.

\subsection{Weak isotropic (IK) cascade}
Incompressible MHD
has two invariants that are separately conserved by nonlinear interactions, namely the energy in each Elsasser mode $E_\pm = <(u^\pm)^2>/2$. 
As a consequence, there are now two energy fluxes, one for each of the two Elsasser energies.
As is visible in eq.~\ref{base1}, nonlinear couplings involve only crossed terms $z^+z^-$, which implies that the nonlinear time of a given field relies on the other field's amplitude:
\be
t_\pm(\ell) = \ell/u^\mp = 1/(ku^\mp)
\label{fort1}
\ee
The energy fluxes thus read
\be
F^\pm_k = (u^\pm)^2/t_\pm = k (u^\pm)^2 u^\mp.
\label{flux-fort}
\ee
However, not all possible triads (k,p,q) in eq.~\ref{base2} contribute equally to 
strong coherent cascades corresponding to the flux expression in eq.~\ref{flux-fort}: 
if the vector associated with wavenumber $q$ is not directed perpendicular to $B_0$
(resonance) then the oscillating exponential factor in eq.~\ref{base2}
weakens the nonlinear couplings and, consequently, the resulting cascade.
Thus, a strictly coherent cascade, i.e. one
that is not impaired by non-resonant scrambling only exists in the field-perpendicular plane.

In a search of a description valid in most of the Fourier space,
\cite{Iroshnikov:1963p9274} and \cite{Kraichnan:1965p9279} (IK)
tacitly ignore the strong cascade 
restricted to the perpendicular $k_x \simeq 0$ plane and focus
on the rest of the Fourier space, thus assuming that the dominant cascade process 
is ruled by non-resonant interactions, with the linear Alfv\'en time being shorter than the nonlinear terms.
In this weak, non-resonant regime, nonlinear couplings drive the cascade via uncorrelated small steps,
each step taking a local Alfv\'en time estimated by the isotropized expression 
\be
t_A^{iso} = 1/(kB_0).
\label{iso}
\ee
The energy cascade is assumed to result from many uncorrelated small steps
which lead to a long time scale for energy transfer
\be
t_\star^\pm =t_\pm (t_\pm/t_A^{iso}) = B_0/(k (u^\mp)^2).
\label{or}
\ee
This leads finally to the expressions for the fluxes
\be
F^\pm_k \sim (u^\pm_k)^2/t_\star^\pm \sim k(u^\pm_k)^2 (u^\mp_k)^2/B_0.
\label{fluxor}
\ee

In the particular case $|u^+| \simeq |u^-|$, the scale-invariance of the two fluxes leads to the IK scaling
\be
u^+ \simeq u^- \sim k^{-1/4}
\ee
which corresponds to the energy spectrum scaling as $E(k) \simeq k^{-3/2}$.
More general solutions are possible, in which the scalings of $u^+$ and $u^-$ differ,
satisfying the condition  $k(u^\pm)^2 (u^\mp)^2 = constant$, i.e., with the sum of spectral slopes being equal to $3$.
Such scalings (the standard IK and as well the generalized $E^\pm$ scalings) have been found in closure calculations and in 2D MHD numerical simulations \citep{1983A&A...126...51G,Pouquet:1988dv}.
Note that in these 2D MHD simulations, there is no mean field: the rms magnetic field plays the role of the mean field $B_0$, in particular in eq.~\ref{iso}, $B_0$ is to be replaced by $b_{rms}$:
\be
t_A^{iso} = (k b_{rms})^{-1}
\ee

\subsection{Anisotropic cascade}
When a strong mean field is present, it has been proposed that, at variance with the previous theory, 
the cascade develops only in the perpendicular direction.
There are two regimes of turbulence corresponding to two different perpendicular scalings in two successive inertial ranges, with a weak cascade followed by a strong one (\cite{Goldreich:1995p4882} (GS), \cite{Ng:1997jd}).
The weak cascade holds for modes for which the characteristic \textit{parallel} Alfv\'en time scale (associated with propagating Alfv\'en waves) is smaller than the perpendicular nonlinear time, or $\chi  \ll 1$, where
\be
\chi = t_A/t_{NL} = k_\bot b/(k_\| B_0)
\label{chi}
\ee

If at the larger scales $\chi \ll 1$, decorrelation of counter-propagating wave packets on
the Alfv\'en time scale will prevent the perpendicular nonlinear couplings to proceed
coherently so that, as in the IK phenomenology, one will obtain a long time scale.
However, the resulting spectral slope differs from the isotropic IK slope because the Alfv\'en time does not vary with $k_\bot$ in this weak turbulence regime, since it is based on the initial parallel scale $\ell_\|^0 = 1/k_\|^0$:
\be
t_A^0 = \ell_\|^0/B_0 = (k_\|^0 B_0)^{-1}
\label{alfven0}
\ee

Replacing the isotropized Alfv\'en time with eq.~\ref{alfven0} in the expression of the energy transfer time (eq~\ref{or}), one gets the scaling
law $u \sim k_\bot^{-1/2}$, and hence the spectral scaling $E(k) \sim k_\bot^{-2}$.

At smaller scales, the weak cascade transforms into a strong one 
when the nonlinear time decreases so as to become 
comparable or smaller than the Alfv\'en time (eq.~\ref{alfven0}) so that $\chi \simeq 1$.
In this strong-turbulence regime, the cascade proceeds as without magnetic field, i.e., the perpendicular scaling is the same as in Kolmogorov phenomenology: the energy spectrum scales as $k_\bot^{-5/3}$.

In the GS picture the strong perpendicular cascade is not completely confined within the perpendicular plane at $k_\|=0$: the 3D spectrum widens in the parallel direction ($k_\| \ne 0$) during the cascade, which does not however correspond to a parallel cascade properly speaking, and so leads to no specific scaling when looking at the 3D spectrum in the parallel direction.
The parallel width of the spectrum results from the ``critical balance'' (CB) between the correlation time of eddies with scale $1/k_\bot$ and the time taken by Alfv\'en waves generated by such eddies to travel a distance $1/k_\|$:
\be
t_A = (k_\| B_0)^{-1}
\ee
The correlation time, in the case of a strong Kolmogorov-like cascade with scaling $u = u_0 (k_\bot/k_0)^{-1/3}$, is given by the nonlinear time $1/(k_\bot u)$:
\be
t_{cor}(k_\bot) = 1/(k_\bot u) = 1/(k_0u_0) \ (k_\bot/k_0)^{-2/3}.
\ee
The critical balance between correlation time and Alfv\'en time then reads $t_{cor} = 1/(k_\| B_0)$ or (identifying $b_{rms}$ with $u_0$):
\be
k_\|/k_0 = (k_\bot/k_0)^{2/3} \ (b_{rms}/B_0).
\label{CB0}
\ee
The $-5/3$ spectral index and the critical balance relation between perpendicular and parallel times has been checked in numerical simulations (\cite{Cho:2002p5358}), and most recently using a reduced MHD shell model by \cite{Verdini:2012hu}.

\subsection{Anisotropic cascade with small-scale cross-helicity}
To explain the occurence of the $k^{-3/2}$ scaling of the energy spectrum in the numerical simulations
by \cite{Muller:2005p705} and the equivalent $k_\|^{-2}$, $k_\bot^{-3/2}$ scalings obtained via SF in \cite{Muller:2003p809},  \cite{2005ApJ...626L..37B, Boldyrev:2006p4917} proposed a variant of the strong perpendicular cascade of GS.

To obtain the $k^{-3/2}$ spectral scaling, one starts with the expression of the energy flux for the strong perpendicular cascade 
 in eq.~\ref{flux-fort},
without the assumption that the two Elsasser species are comparable. 
Instead, we assume that the dynamics generates spontaneously different scaling laws for a dominant Elsasser species (say $u^+$) and for a subdominant species (say $u^-$).
The idea (confirmed by the numerical simulations of \cite{2006PhRvL..97y5002M}) is (i) that the largest scales show
only small $u^+/u^-$ imbalance; (ii) that the imbalance increases with wavenumber; (iii) that the dominant species is with equal probability either $u^+$ or $u^-$.
To obtain the $k^{-3/2}$ spectral scaling, or $u^+ \propto k^{-1/4}$, it is required that $u^+/u^- \propto k^{1/4}$.
In that case, we have from eq.~\ref{flux-fort}:
\be
F^+ = k (u^+)^2 u^- = k (u^+)^3 k^{-1/4} = k^{3/4}(u^+)^3
\ee
which indeed leads to the required IK scaling $u^+ \propto k^{-1/4}$ when assuming $F^+ = constant$.

Several remarks are in order. 
First, the version just given differs from the original one. The original version is formulated in terms of the velocity-magnetic field alignment, and measures the scaling law  of the angle $\alpha$ (\cite{2006PhRvL..97y5002M}):
\be
\sin \alpha = <|\delta u(\mathbf{L}) \times \delta B(\mathbf{L}) |> / <|\delta u(\mathbf{L}) || \delta B(\mathbf{L}) |> 
\label{delta}
\ee
where $\delta B_i(\mathbf{L}) = B_i(\mathbf{x}+\mathbf{L}) - B_i(\mathbf{x})$ (and a similar expression for $\delta u$), and the brackets denote spatial averaging on the position $x$.
The two versions (either using the angle $\alpha$ or the $u^-/u^+$ ratio) 
are actually equivalent, inasmuch as, assuming equipartition of velocity and magnetic energies $|u| \simeq |b|$ at small scales, one has $1 - (u^-/u^+)^2 \propto \cos \alpha$, which implies for small $u^-/u^+$:
\be
\alpha \simeq u^-/u^+
\label{ssda1}
\ee

The scaling 
\be
\alpha \sim k^{-1/4}
\label{quart}
\ee
has been found to hold in 3D MHD simulations by \cite{2006PhRvL..97y5002M}. In these simulations, one has $|u^+| \simeq |u^-|$ at the largest scales, 
while $u^- \simeq u^+ k^{-1/4} \simeq k^{-1/2}$: in other words, the velocity and magnetic field 
fluctuations become completely aligned at small enough scales. 
We will call in the following this process ``small scale dynamic alignment'' (SSDA).
This is to be compared with the regime of ``large scale dynamic alignment'' studied by 
\cite{1983A&A...126...51G} and \cite{Pouquet:1988dv} in which, on the contrary,
the alignment lies at large scales and goes to zero at small scales.

Actually, these different scalings for $u^+$ and $u^-$ are in contradiction with the basic 
strong (i.e., resonant) expressions for the energy flux (eq.~\ref{flux-fort}).
Indeed, the invariance of the two fluxes during the cascade implies a single solution:
the two scalings must both be $u^+ \propto u^- \propto k^{-1/3}$ and nothing else
\citep{2003ApJ...582.1220L,2008ApJ...682.1070B}.
The only way out of this paradox consists in assuming that in eq.~\ref{flux-fort} the dimensional expression for the
minor flux $F^-$ is \textit{not} valid.

\section{Re-analyzing the GM10 regime}
The turbulent regime reported by \cite{Muller:2005p705} has a mean field $B_0 = 5 b_{rms}$.
The corresponding MHD equations include given stationary large scale flows $U_1$ and magnetic fields $B_1$ with wavenumbers satisfying $1 \le k \le 2$, and equal energies; the rest of the spectrum evolves freely, so
that the equations read:
\begin{align}
\partial_t \textbf z^\pm 
\pm (\textbf B_0 \cdot \nabla) \textbf z^\pm
+(\textbf z^\mp \cdot \nabla) \textbf z^\pm \nonumber + \nabla P \\  
+(\textbf U_1 \cdot \nabla) \textbf z^\pm
\pm (\textbf B_1 \cdot \nabla) \textbf z^\pm \nonumber \\
+(\textbf z^\mp \cdot \nabla) \textbf U_1 
\mp (\textbf z^\mp \cdot \nabla) \textbf B_1  
 = 0
\label{base2a}
\end{align}
with the fluctuating quantities $z^\pm$ having wavenumbers $k > 2$.
As the forcing is isotropic with $B_0=5 b_{rms}$, 
the cascade should in principle follow a weak regime in the largest scales at least, 
since (eq.~\ref{chi}) $\chi \simeq 1/5$
at the largest scales.
The aspect ratio of the simulation domain is unity, which is not of concern in the case of weak turbulence
(we come back on this point in the discussion).
Hence, with our configuration with weak coupling at large scales we should obtain in principle a purely perpendicular cascade
with a fixed parallel extent of the spectrum until the coupling becomes strong and thus ruled by the 
critical balance (see e.g., \citep{Verdini:2012hu}).
However, this scenario is not found here.

Indeed, the analysis by \cite{2010PhRvE..82b6406G}, shows the following properties of the 3D spectrum: 
(i) the scaling is $k^{-3/2}$ in all directions; in other words, instead of being mainly perpendicular, 
it extends in all directions; 
(ii) the $-3/2$ power-law range extends largely into the weak $\chi \ll 1$ domain.
These properties make the cascade resemble the IK cascade, with however a substantial angular variation of the amplitude with angle $\theta$ between wave vector and mean field, such that the respective sizes of the perpendicular and parallel power-law ranges are in a ratio close to $B_0/b_{rms}=5$.

\subsection{Diagnostic tools}
We first recall standard definitions.
The 1D reduced spectra (SP) vs parallel ($k_\|$) and perpendicular ($k_\bot$) wavenumbers are obtained by integrating the 3D spectra on planes perpendicular to the chosen direction:
\be
E(k_\|)= \int \int E_3(k_\|,k_y,k_z) dk_y dk_z
\label{E1Dpar}
\ee
and
\be
E(k_\bot)= \int \int E_3(k_x,k_\bot,k_z) dk_x dk_z
\label{E1Dper}
\ee
Structure functions (SF) are another way to measure scaling laws in turbulent systems.
We will use second-order SF built on the magnetic field fluctuations:
\be
SF(\mathbf{L}) = <\delta B^2(\mathbf{L})>.
\label{SF1}
\ee
One considers mainly SF vs $L_\|$ and $L_\bot$, where parallel and perpendicular refer to the mean field direction. The mean field may be computed globally or locally: in the latter case, the average is replaced by a discrete mean (method 1 of \cite{Cho:2000p6223}).

Global SF (that is, referring to global mean field $B_0$) and reduced spectra (both parallel and perpendicular) are related by the following equation:
\begin{eqnarray}
SF(L) = 2 [E - \mathcal{F}^{-1}E(k)]
\label{SPSF}
\end{eqnarray}
where $E$ is the total magnetic energy and $\mathcal{F}^{-1}$ denotes the inverse Fourier transform with respect to the variable $k$.

When power-laws are present, the correspondence between SP and global SF slopes is obtained by using the following classical relation:
\be
SF(L) = \delta b^2(L) \sim k E(k)
\label{dim}
\ee
with $k \simeq 1/L$.
This is the relation that is universally used to express the SF scaling laws in terms of SP scaling laws, the latter being more familiar. However, in our case, this relation is broken due to the importance of non-scaling structures in Fourier space (see below and Appendix).

\subsection{3D Spectral vs SF structure \label{anifou}}
\begin{figure}
\begin{center}
\includegraphics [width=\linewidth]{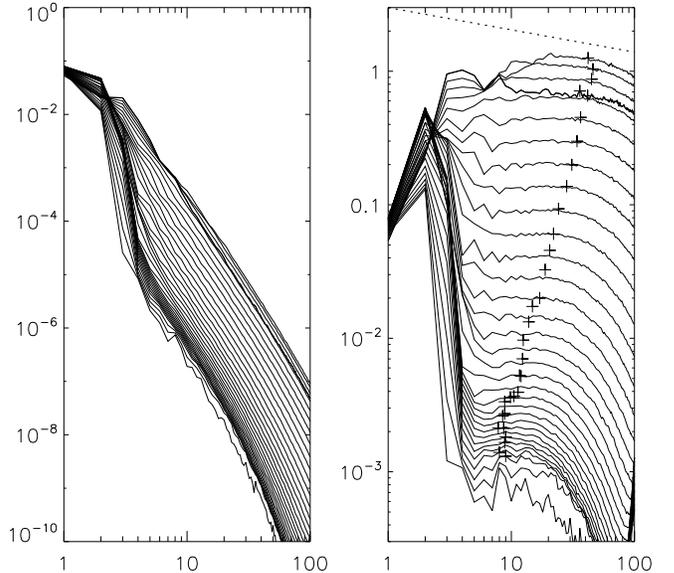}
\caption{
The 3D GM10 spectrum.
3D GM10 spectrum $E_3(k_\|,k_\bot)$ 
averaged about the azimuthal angle with respect to $B_0$, sum of the kinetic and magnetic spectrum. On represents radial cuts at different angles $\theta$ with respect to the mean field $B_0$,
vs wavenumber $k$, with the angle $\theta$ between wave vector and mean field $B_0$ varying from $\pi/2$ (top bold curve) to $0$.
Left: original curves; Right: same curves, compensated by $k^{-3/2-2}$. Dotted line: $k^{-5/3-2}$ scaling. ``+'' symbols: position of half the dissipative wavenumber at each angle $\theta$.
}
\label{fig1}
\end{center}
\end{figure}

We show in fig.~\ref{fig1} radial cuts of the 3D spectrum, obtained after averaging over the azimuthal angle $\phi$ around the mean field $B_0$. 
It shows radial cuts $E_3(k,\theta)$ vs wavenumber $k$ for a series of angles $\theta=\angle (k,B_0)$, as in fig.3a of GM10. In the left panel, the original spectra are shown; in the right panel the profiles
are compensated by $k^{-3/2-2}$, to highlight deviations of the slope with respect to the average 1D $-3/2$ slope, the dotted curve indicating the $k^{-5/3-2}$ scaling.
(Note the difference of $-2$ between the 3D and 1D spectral indices).
Formally, the algebraic form that fits this structure is:
\be
E_3(k,\theta) = A(\theta) k^{-m-2} 
\label{GM}
\ee
with a scaling index $m=3/2$, independent of the angle $\theta$.

We also indicate for each angle the half dissipative wavenumber $k_d(\theta)/2$ by plotting a plus sign on the corresponding curve, for later use in the Appendix. 
The dissipative wavenumber $k_d(\theta)$ is obtained as in GM10 by determining the maximum of each radial profile of the 3D spectrum weighted by $k^4$ at each angle $\theta$.

Determining the spectral slope requires discarding two non-scaling wavenumber ranges: (i) the dissipative range with $k>k_d/2$ and (ii) the forcing range.
The forcing range ($1\le k \le 8$) includes the truly forced scales ($1 \le k \le 2$) and the intermediate scales that make the transition with the power-law range (see fig.~4 in GM10).

In doing so, one obtains a radial slope which is in average close to $-3/2-2$ (eq.~\ref{GM}),
with some exceptions at larger angles (see the dotted line that indicates the $-5/3-2$ slope).
Note that the slope can be determined with sufficient accuracy only for directions with angle larger than $\theta \simeq 14^o$: for smaller angles, the signal is too noisy and the power-law range too short.

On the other hand, SF exhibit anisotropic scalings (\cite{Muller:2003p809}): close to $L_\bot^{1/2}$ and $L_\|$.

One can in principle explain why the measured spectrum has a much more isotropic scaling than the SF.
The point is that SF can be measured in a frame attached to the local mean field, not the 3D spectrum that has to be mesured in the global frame attached to the global mean field.
In the global frame, the
clear dominance of the perpendicular direction that appears in the frame attached to the local mean field
is smoothed out, being distributed to a large interval of oblique directions, due to the random wandering of the local mean field with respect to the global mean field direction.

If this argument is correct, then SF parallel and perpendicular scalings should collapse, leading to a single scaling close to the perpendicular one, when considering parallel and perpendicular directions defined with respect to the global mean field.
To check this point, we have computed the \textit{global} and  \textit{local} SF. 
\begin{figure}[t]
\begin{center}
\includegraphics [width=0.7\linewidth]{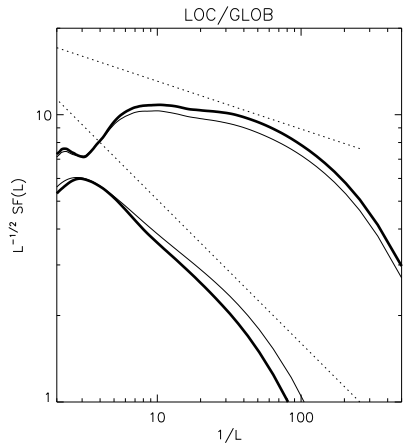}
\caption{
GM10 simulation. Local (thick lines) and global (thin lines) second order structure functions SF vs parallel and perpendicular separations $L$, for the magnetic field fluctuations, compensated by $L^{1/2}$. 
Upper curves: perpendicular SF; bottom curves: parallel SF.
Note that plateaux are in principle equivalent to reduced 1D spectra scaling as $k^{-3/2}$.
Dotted lines indicate the $L^{2/3}$ and $L$ scalings (in principle equivalent to, respectively, 
$k^{-5/3}$ and $k^{-2}$ 1D reduced spectral scaling).
}
\label{figSFgloloc}
\end{center}
\end{figure}

The result is shown in Fig.~\ref{figSFgloloc}. The two SF pairs (parallel and perpendicular) are plotted, the \textit{local} SF with thick lines and the \textit{global} SF with thin lines. 
All SF are compensated by a $L^{1/2}$ scaling, which corresponds in principle to a $k^{-3/2}$ scaling for a 1D spectrum (eq.~\ref{dim}).
The dotted lines show the $L^{2/3}$ and $L$ scalings, corresponding respectively to $k_\|^{-5/3}$ and $k_\bot^{-2}$ spectral scalings.

Both the local and global SF show scalings close to, respectively, the IK prediction in the perpendicular direction and the critical balance prediction for the parallel direction:
the local SF scale as ($L_\bot^{0.55}, L_\|$), and the global SF as  $L_\bot^{0.58}, L_\|^{0.9}$.
This confirms the earlier results by \cite{Muller:2003p809}, and at the same time shows
that in the present case, there is actually no significant difference between measuring SF in the local and global frames.

We conclude that the spectral properties considered above are not the plain consequence of the smoothing of the local anisotropic dynamics due to the random fluctuations of the local mean field.
Thus, the amplitude anisotropy of the spectrum and the scaling anisotropy of the SF are two complementary aspects of the GM10 turbulent regime.
The origin of the failure of the classical eq.~\ref{dim} that usually relates scaling in Fourier and real space lies in the non-scaling character of the large isotropic scales in GM10 (see Appendix).

\begin{figure}
\begin{center}
\includegraphics [width=0.69\linewidth]{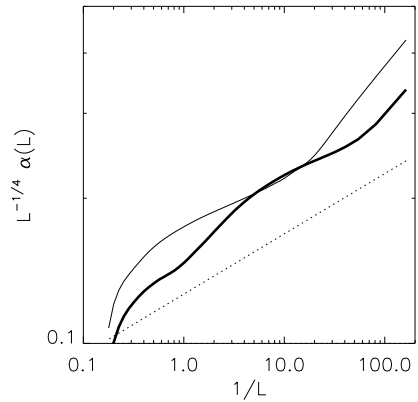}
\caption{
Scaling of angle $\alpha$ between magnetic and velocity fluctuations between points separated by a distance $L$ in the GM10 data, vs $1/L$.
thick curve: the vector $\mathbf{L}$ is perpendicular to the mean field direction;
solid curve: the vector $\mathbf{L}$ is parallel to the mean field direction. 
The profiles are compensated by $L^{1/4}$.
Dotted line indicates $L^{1/8}$ scaling for reference.
}
\label{figSSDA}
\end{center}
\end{figure}

\subsection{Small scale dynamical alignment}

In order to determine what could be the correct cascade scenario at the origin of the measured perpendicular $k_\bot^{-3/2}$ spectral scaling, either the IK or the SSDA scenario, we have examined
how the angle $\alpha$ between velocity and magnetic field fluctuations depends on scale $L$.
In the SSDA scenario, the $k_\bot^{-3/2}$ energy spectral scaling (or, equivalently, the $u \sim k_\bot^{-1/4}$ amplitude scaling) results from the angle scaling as $\alpha \sim k_\bot^{-1/4}$ (eq.~\ref{quart}, see \cite{2006PhRvL..97y5002M}).

Fig.~\ref{figSSDA} gives the measured $\alpha$ vs $1/L$ averaged over several successive times, 
compensated by $L^{1/4}$, separately for the vector $\mathbf{L}$ being in a direction perpendicular to $B_0$ (thick line), or the vector $\mathbf{L}$ being in a direction parallel to the mean field (thin line).
The resulting scaling is in both directions at best $\alpha \sim L^{1/8}$.
Expressed in terms of the amplitude ratio $u^-/u^+$ (subdominant over dominant amplitude) this reads (eq.~\ref{ssda1}):
\be
u^-/u^+ \sim k^{-\beta} \sim k^{-1/8}
\label{angle}
\ee
We thus conclude that 
the SSDA cannot, alone, be at the origin of the GM10 regime. 

\subsection{Transition from the isotropic to the perpendicular cascade}
A way to pass from the isotropic GM10 regime to the more familiar purely perpendicular cascade 
consists in relaxing the forcing (we will see another way in the discussion).
We switch off the forcing after $48.3$ nonlinear times and follow the evolution up to time $t=60.9$. 
We define five subintervals of time, denoted by roman numerals, from I to V. Interval I denotes the interval of time during which the properties of the forced GM10 regime have been measured, covering ten outputs from time $46.8$ up to $49.5$.
Intervals II , III, IV and V denote intervals which cover the decay phase, containing the following output times: $[50.4,51.3,52.2]$ (II), $[53.1,54.0,54.9]$ (III), $[55.8,56.5,57.4]$ (IV), $[58.3,59.2,60.1,60.9]$ (V).
\begin{figure}
\begin{center}
\includegraphics [width=\linewidth]{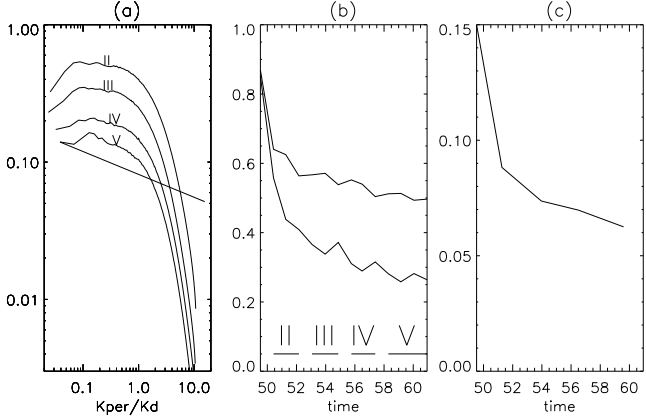}
\caption{
Evolution, during ten nonlinear times, of GM10 turbulence when the equipartition condition on large scales is relaxed (followed).
(a) Reduced spectra $E(k_\bot)$ of total energy.
Each of the four spectra (marked II, III, IV, and V) is averaged over three outputs (except period V for which there are four outputs) which cover a period of about two and a half nonlinear times.
Dotted line: $k^{-5/3}$ spectrum.
(b) Rms amplitudes of magnetic (thick line) and kinetic fluctuations (thin line) vs time starting with GM10 regime. Rms amplitudes are computed in the forcing scale range $1 \le k \le 2$. The four periods II, III, IV and V are reported on the abscissa.
(c) Decay of aspect ratio of the total energy contours (see text for definition).
}
\label{figfilm}
\end{center}
\end{figure}

\begin{figure}
\begin{center}
\includegraphics [width=\linewidth]{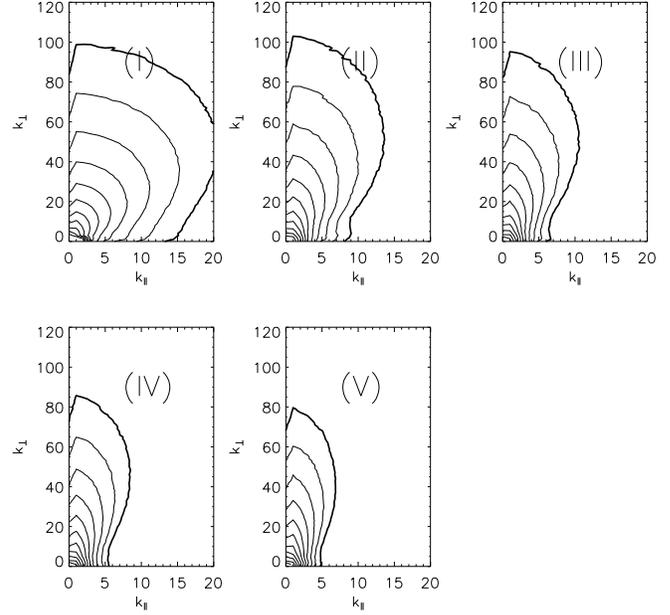}
\caption{
Evolution, during ten nonlinear times, of the 3D spectrum $E_3(k_\|,k_\bot)$ when the equipartition condition on large scales is relaxed.
(I) Forced GM10 regime;
(II-III-IV-V): Decaying phases. Each 3D spectrum is averaged over three outputs which cover the corresponding periods, except for phase V which contains four outputs. 
}
\label{figfilm2}
\end{center}
\end{figure}

\begin{figure}
\begin{center}
\includegraphics [width=\linewidth]{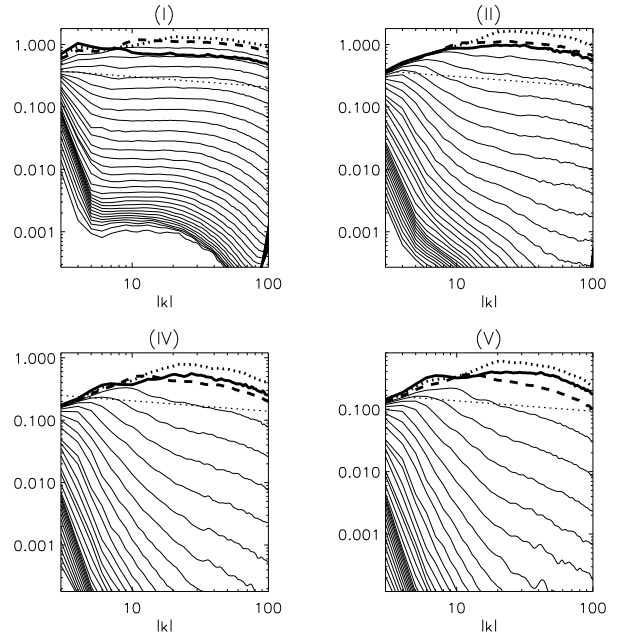}
\caption{
Evolution, during ten nonlinear times, of the family of radial cuts of the 3D spectrum $E_3(k,\theta)$ when the equipartition condition on large scales is relaxed, averaged during different periods.
(I) Forced GM10 regime;
(II-IV-V): Decaying phases. Each family of radial cuts is averaged over three outputs which cover the corresponding periods, except for period V which contains four outputs. Period III is omitted.
}
\label{figfilm3}
\end{center}
\end{figure}

As we know, the forcing regime consists in imposing velocity and magnetic field fluctuations with wavenumbers between ($1\le k \le 2$). One observes during the decay period (intervals II to V) that:
\begin{enumerate}
\item the perpendicular spectral scaling passes progressively from $k^{-3/2}$
to $k^{-5/3}$ (fig.~\ref{figfilm}a); 
\item the previously forced scales ($1 \le k \le 2$) rapidly pass from initial equipartition with $u_{rms} \simeq b_{rms}\simeq1$ to a dominance of magnetic energy (fig.~\ref{figfilm}b); 
\item spectral anisotropy increases, i.e., the parallel and oblique directions show less and less excitation compared to the perpendicular direction (fig.~\ref{figfilm}c and fig.~\ref{figfilm2}).
\end{enumerate}

The anisotropy shown in fig.~\ref{figfilm}c is measured by plotting the ratio between the parallel ($k_{\|\star}$) and perpendicular ($k_{\bot\star}$) extents of a given level of isocontour, that corresponding approximately to the dissipative isocontour during the forced regime.
These two wavenumbers are given by the intersection of the specific isocontour with the $k_\|$ and the $k_\bot$ axes.
The decay of the parallel extent of the spectrum compared to the perpendicular one (fig.~\ref{figfilm}c) is approximately following the decay of the ratio $b_{rms}/B_0$ with time (fig.~\ref{figfilm}b). 
This anisotropy variation is a quantitative measure of the evolution shown by plotting the successive spectral isocontours averaged during the five time intervals I to V, in fig.~\ref{figfilm2}.

This evolution of the spectral anisotropy actually hides a much deeper modification of the 3D spectrum, namely the complete loss of any scaling except in the perpendicular direction.
This is revealed clearly in fig.~\ref{figfilm3}, where we show the radial cuts of the spectra compensated by the $k^{-3/2-2}$ law,
averaged for the periods I, II, III and V (period IV is not shown). 
Note that in fig.~\ref{figfilm3}, we have used a two-point smoothing along each radial ray (not present e.g. in fig.~\ref{fig1}b), in order to make clearer the initial $k^{-3/2-2}$ scaling of the forced regime, together with the disappearance of scaling during the decaying phase.

In conclusion, relaxing the freezing of the scale range $1 \le k \le 2$ leads to the loss of the isotropic $k^{-1.5}$ specific 
of the GM10 regime, and ends up with a pure perpendicular cascade, with a clear-cut $k^{-5/3}$ scaling for the reduced perpendicular spectrum, and a large dominance of the magnetic fluctuations in the previously forced range.
It is interesting to note that a similar transition to the same regime can be obtained by maintaining frozen large scales $1 \le k \le 2$, but replacing the magnetic-kinetic equipartition by a large dominance of the magnetic energy over kinetic energy.
Note that we do not show any figure as they are completely equivalent to those already discussed.

The possible occurrence of the GM10 regime in astrophysical flows will be discussed in the last Section, as well as its status compared to that of the classical perpendicular cascade with 5/3 scaling law.

\section{A new model for 3D MHD turbulence}\label{model}
We propose now a model that describes how a mainly perpendicular cascade may be transformed into a cascade filling the Fourier space in all directions.
In the following, we will distinguish the two fields $u^+$ and $u^-$, with the former field denoting the \textit{dominant} field (when assuming that some SSDA is present). However, the argument does not really require SSDA to be active, i.e., the cross-helicity scaling may as well be assumed to be zero.

\subsection{Perpendicular cascade and critical balance extension}
Since the ($k_\|=0$) plane perpendicular to the mean field is energetically dominant,
we first consider a scenario for the perpendicular scaling.
We assume that the perpendicular cascade is the 2D IK cascade, controlled by $b_{rms}$ evaluated in the perpendicular plane, namely with the characteristic Alfv\'en time $(k_\perp b_{rms})^{-1}$.
The energy fluxes thus read (cf. eq.~\ref{fluxor}):
\be
F^\pm_k \sim (u^\pm_k)^2/t_\star^\pm \sim k(u^\pm_k)^2 (u^\mp_k)^2/b_{rms}
\label{fluxorbrms}
\ee
In the case with $|z^+| \simeq |z^-|$, this reduces to $F \sim ku^4$, and the flux invariance leads to 
the usual IK scaling $u \simeq k^{-1/4}$, 
i.e., an energy spectrum as $E_k \simeq k^{-3/2}$.

More generally, if we take into account the non-zero cross-helicity scaling (eq.~\ref{angle}), we find (using constancy of the flux $F^+$) that
$u^+_k = k^{-1/4+\beta/2}$ or
$E^+_k = k^{-3/2+\beta}$.
In conclusion, the spectral slope is $m=-3/2$ when neglecting the cross-helicity, and $m=-1.375$ if we take the rough approximation $\beta = 1/8$.


Assuming seeds of fluctuations with small $k_\|$ are present, the spectrum resulting
from the previous 2D IK perpendicular cascade will have a limited parallel extent.
We assume this parallel extent $k_\|$ is controlled by the critical balance at each scale $1/k_\bot$ between the Alfv\'en time computed on $k_\|$ and the correlation time, which
is the shortest available characteristic time produced by the perpendicular cascade. 
While in the usual strong cascade, the correlation time is given by the nonlinear time, 
here, in contrast, it is given by the \textit{perpendicular} Alfv\'en time, namely $1/(k_\bot b_{rms})$.
This leads to a parallel spectral extent $k_\|$ scaling as 
\be
k_\| B_0 \simeq k_\bot b_{rms}.
\label{cbik}
\ee
This will be used below.

\subsection{Oblique cascade}
The 2D IK cascade is still quasi-perpendicular.
We now propose a scenario that uses the 2D IK perpendicular cascade as a driver to generate an oblique cascade. At the end, this will lead to \textit{scaling isotropy}, i.e., isotropy of the spectral index.

\subsubsection{Quasi-resonant cascade}
We start with the basic eq.~\ref{base2} and retain only $(k,p,q)$ triads such that the 
coupling term is quasi-resonant.
This means that the term $e^{2iq_\| B_0 t}$ should oscillate on a time scale longer than
the nonlinear time $t_+$ of the main Elsasser species $u^+$:
\begin{equation}
q_\| B_0 \le 1/t_+ \simeq qu^-_q
\label{BC2}
\end{equation}
The \textit{largest} value of $q_\|$ compatible with quasi-resonance is:
\begin{equation}
q_\|^{max} \simeq qu^-_q/B_0
\label{BC3}
\end{equation}
Note that vectors $q$ satisfying eq.~\ref{BC3} are to be found inside the critical balance cone (eq.~\ref{cbik})
$q_\| = q b_{rms}/B_0$ as $u^-_q / b_{rms} \rightarrow 0$ when $q$ increases. 
This means that the perpendicular cascade alone provides  the required reservoir of quasi-perpendicular modes $u^-_q$.

The formal evolution eq.~\ref{base2} becomes now for the dominant field $u^+$:
\begin{align}
\partial_t \widehat{u^+_k} &=& k \int d^3q \ \widehat{u^+_p} \widehat{u^-_q} e^{\mp 2i q.B_0 t} 
\nonumber \\
&\simeq&  k \int_{QR} d^3q  \ \widehat{u^+_p} \widehat{u^-_q}
\label{baseQR}
\end{align}
where the mention $QR$ indicates that the sum deals with only the vectors $q$ that satisfy the quasi-resonance condition eq.~\ref{BC3}.

Using the localness hypothesis (eq.~\ref{localness}) allows to estimate
how the nonlinear time scale is modified by the quasi-resonance constraint.
While in the full strong cascade with $B_0=0$ the inverse characteristic nonlinear time can be estimated as in eq.~\ref{fort1} $(t^+)^{-1} \sim k \int d^3q \ \widehat{u^-_q} \sim k u^-_q$, we obtain now,
\be
(t^+)^{-1} \sim R k \int d^3q \ \widehat{u^-_q} \sim R k u^-_q
\label{tstarR}
\ee 
where $R$ denotes the proportion of active triads remaining due to the quasi-resonance constraint.
The reduction factor $R$ will be estimated below. We first describe in some detail 
the cascade process.

\subsubsection{The ricochet process}
To fill the rest of the $k_\|,k_\bot$ plane outside the perpendicular domain (and its critical balance extension), we propose now a specific model, that uses quasi-resonant triads in the sense given just above, i.e., with the third wavevector $q$ being quasi-perpendicular.

\begin{figure}
\includegraphics [width=\linewidth]{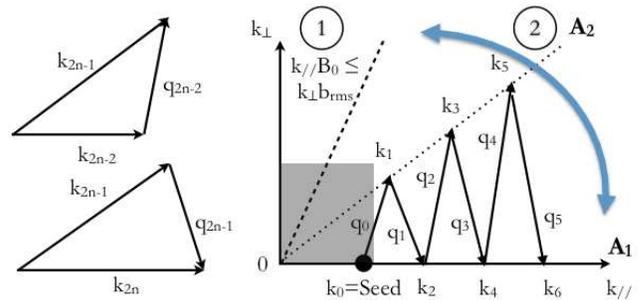}
\caption{
Two-step scenario leading to an isotropic scaling in MHD turbulence with large mean field.
Left panel: elementary triads used to propagate excitation towards small scales along the two rays $A_1$ and $A_2$ in the right panel.
Right panel: propagation of excitation along a pair of oblique lines $A_1$ and $A_2$.
(1) region of perpendicular IK cascade with critical balance extension (dashed line, eq.~\ref{cbik}).
(2) region of coupled oblique cascades or ``ricochet process'' along the oblique lines (here $A_1, A_2$), driven by quasi-perpendicular modes $q_i$.
The process starts with a seed $k=k_0$ outside the critical balance region (1), 
provided by the isotropic forcing, marked in gray.
}
\label{fig3}
\end{figure}

Fig.~\ref{fig3} sketches the model.
The left panel shows two elementary triads, each with one quasi-perpendicular vector $q$.
The right panel shows the $(k_\|, k_\bot)$ plane with a pair of oblique lines $A_1, A_2$ passing through the origin. 
These two oblique lines trace (among many others possible oblique lines) paths along which the excitation
is able to propagate, using a series of elementary triads such as shown in the left panel.
These triads all have their third wavevectors $q$ satisfying the quasi-resonance condition.
The quasi-perpendicular vectors $q$ couple the two lines together, allowing the two radial cascades to proceed together, in parallel.

Note that for clarity one of our two oblique lines ($A_1$) is horizontal, but there is no real necessity for that: any pair of oblique lines (with an initial large-scale seed provided by the forcing) contributes in principle to the propagating the excitation towards small scales, as soon as the angle between them is non-vanishing (as this would correspond to a vanishing contribution to the convolution integral).

We call the process sketched in fig.~\ref{fig3} the ``ricochet'' process, since excitation propagates along the two lines by successive bounces from one line to the other, thanks to the critical balance reservoir of wavevectors $q_i$.

Note that we don't give any argument showing that energy is indeed flowing along these rays.
Detailed tests that this is indeed occurring is outside the scope of this work.
Strictly speaking, our arguments only indicate that this is possible. 
However, we remark that the triads considered are representative of triads able to drive a cascade in 
general: the only selection criterion required is that
(i) they are ``local'', that is, $k \simeq p \simeq q$) (ii) the $q$ vector satisfies quasi-resonance.

We estimate the energy flux $F_{ob}$ in our oblique cascade by dividing as usual the energy $(u_k^+)^2$ at scale $1/k$ by its characteristic time $t^+$ (estimated from the nonlinear term in eq.~\ref{tstarR}):
\be
F_{ob}^+ \simeq (u_k^+)^2 /t^+ \simeq R k (u_k^+)^2 u_q^-
\label{FOB1}
\ee
This is indeed the expression corresponding to a strong cascade (the evaluation of the factor $R$ is to be done below).
Two remarks are in order.
(i) We don't consider separately different flux expression for different oblique directions: we estimate globally the ``oblique'' flux, i.e. the flux from the large scales flowing in directions other than the perpendicular plane. 
(ii) The flux expression is not completely standard: reflecting the structure of the nonlinear kernel, 
it mixes the dominant ``oblique'' field $z_k^+$ with the sub-dominant ``perpendicular'' field $z_q^-$.

\subsubsection{Flux reduction} 
Due to the quasi-resonance constraint (eq.~\ref{BC3}), a number of triads are eliminated compared
to all possible ones that actually would contribute in the zero mean field case.
At a fixed wavenumber $k \simeq q$, only triads with $q_\|/q \le u^-_q/B_0$ contribute to the cascade along the parallel and oblique directions, while
the triads contributing to the cascade in the absence of mean field are characterized by $0 < q_x \lesssim q$.
\begin{figure}
\begin{center}
\includegraphics [width=0.75\linewidth]{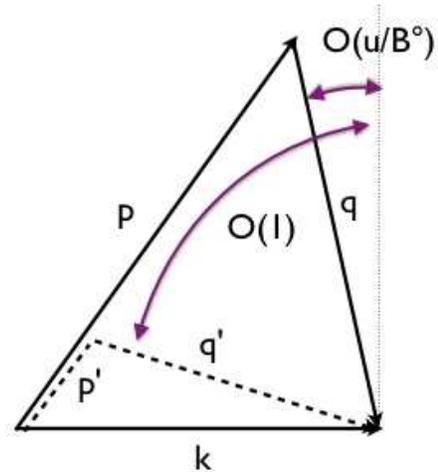}
\caption{
The condition of quasi-resonance which forces the vector $q$ to be quasi-perpendicular reduces the number of interacting triads by a reduction factor of order $R = u/B_0$ (eq.~\ref{eq-R}).
}
\label{fig-QR}
\end{center}
\end{figure}
As sketched in fig.~\ref{fig-QR},
the contributing subset of quasi-resonant triads is thus only a fraction of the total number of triads, of order $R$:
\be
R \simeq q_\|^{max}/q \simeq u^-_q/B_0
\label{eq-R}
\ee
Accordingly, the energy flux in the oblique direction (eq.~\ref{FOB1}) becomes: 
\be
F_{ob}^+ \simeq k (u_k^+)^2 u_q^- R \sim k (u^+_k)^2 (u^-_q)^2/B_0 
\label{FLFL}
\ee
Comparing eqs~\ref{fluxorbrms} (with $k$ replaced by $q$) and \ref{FLFL}
shows that the resulting scaling laws vs $q$ and $k$ are identical.

The spectral extent, which is controlled by viscosity, provides a test of our phenomenology of coupled perpendicular and oblique cascades. 
Indeed, at the dissipative scale $1/k_d$, one should find equal transfer and dissipative times. 
This leads to two equalities close to one another, depending on whether we consider respectively the parallel or perpendicular cascade:
\begin{eqnarray}
 \nu k^2 \simeq k (u_q^-)^2 / B_0
 \\
 \nu k^2  \simeq k (u_q^-)^2 /b_{rms} 
\end{eqnarray}
hence
\be
k_{d_\parallel}/k_{d_\bot} = b_{rms}/B_0
\ee
This relation has been found numerically by GM10.

\section{Conclusion}

The present cascade scenario proposed here is motivated by the 3D spectrum observed in GM10.
The GM10 regime may be called a three-dimensional Iroshnikov-Kraichnan regime, having the following characteristic properties:
(i) large-scale energetic equipartition (instead of magnetic dominance);
(ii) 3/2 scaling (instead of 5/3);
(iii) isotropic scaling (instead of perpendicular).

We propose here a new turbulent cascade mechanism to explain it, which is a combination of
the weak Iroshnikov-Kraichnan dynamics governing energy transfer in the 
field-perpendicular plane and the ricochet process distributing energy
quasi-resonantly along all other directions.  This yields turbulence
properties consistent with the main findings of GM10: (i) an
inertial-range power law exponent independent of direction, (ii) a
direction-dependent inertial range extent $\sim
b_{rms}/B_0$. This process asymptotically approaches the 2D IK-cascade
as $B_0$ increases.

The new transfer mechanism is different from the commonly accepted 
resonant weak-turbulence cascade as well as from the 
critically balanced strong turbulence cascade, both resulting in strictly perpendicular 
energy transfer. 
The new process generates the significant 
part of the observed energy distribution in the oblique and parallel directions,
which is not explicable by current strong turbulence phenomenologies based on critical balance 
(GS).

The whole analysis is rendered more complex since contrary to the energy
spectrum the structure functions display anisotropic scaling. By
additional numerical experiments descibed in the Appendix this is
shown to be the consequence of the presence of a transition region
between forced and inertial scales which breaks the simple relation
between spectral and structure-function scaling.

Solar wind turbulence indirectly shows a similar apparent contradiction
between SF and spectral scalings.  There, SF scale as predicted by the
critical balance, when measured with respect to the local mean field
\citep{2008PhRvL.101q5005H}.  However, one cannot decide unambiguously
whether the $k_\|^{-2}$ scaling revealed by the SF is associated with
an indirect contribution of the perpendicular cascade as in the
critical balance theory, or is the signature of a true parallel
cascade \citep{Horbury:2011dh}.  

\begin{figure}
\begin{center}
\includegraphics [width=\linewidth]{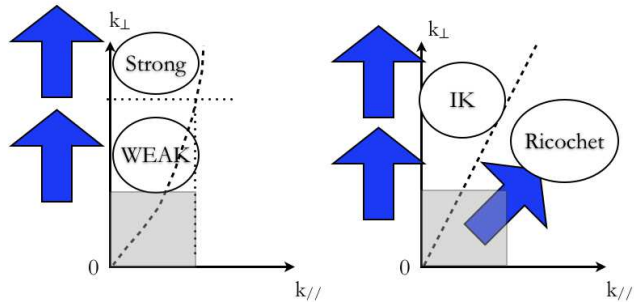}
\caption{
The two possible regimes with mean field when forcing isotropically the large scales.
Left: the standard perpendicular cascade bounded by critical balance (eq.~\ref{CB0}, bold dashed line), with weak cascade at large scale followed (above the horizontal dotted line) by a strong cascade at smaller scales.
Right: the GM10 cascade made of a weak IK perpendicular cascade bounded by the IK version of critical balance, (see eq.~\ref{cbik}, bold dashed line), and the oblique ricochet cascade constrained by the quasi-resonant condition (eq.~\ref{BC2}).
The gray region is the forced area.}
\label{fig-der}
\end{center}
\end{figure}

Fig.~\ref{fig-der} sketches the basic characteristics of the two scenarios of turbulent cascades in Fourier space.  
The large-scale conditions are the same, with isotropically forced scales (represented by a gray square).  
The classical perpendicular cascade scenario, sketched in the left
panel, is governed by the critical balance (with possible
small-scale cross-helicity scaling); the new scenario, sketched in the
right panel, is based on the weak IK perpendicular cascade, without the critical balance playing a visible role.  
While in the classical process, the parallel spectral extent
remains first constant and then follows the critical balance rule, 
the ricochet process immediately starts at largest freely-evolving scales
to extend the spectrum in all oblique and parallel directions, due to the
possibility of quasi-resonant couplings with the slow perpendicular cascade.

Our results from section 3.4 show that, among the two regimes, the classical $5/3$ regime is the most robust,
when no forcing is imposed on the system.
However, if magnetic-kinetic equipartition is imposed on the large scales, then the 3D IK regime of GM10 develops.

A coexistence between the two regimes seems to be observed in the solar wind, where the spectral slope varies systematically with the wind speed, with correspondingly the large-scale magnetic excess: slopes close to $5/3$ are observed in slow winds where a large magnetic excess is observed, while shallower spectra with slopes approaching $3/2$ are observed in fast winds with no magnetic excess (\cite{1991AnGeo...9..416G} and work in progress).

How the large-scale magnetic excess can control the cascade remains to be explained. Two points are interesting to note in this regard and should guide the research. First, a similar relation between magnetic excess and spectral scaling has been recently found as well in MHD turbulence with no mean field \cite{2010PhRvE..81a6318L,2014PhRvE..89d3017K}). 
Second, the reverse phenomenon, namely how the cascade controls the magnetic excess, has been clearly shown to occur in MHD turbulence with no mean field as well as with mean field (in the GM10 regime) \citep{Muller:2005p705}:
the magnetic excess results from the competition between a local dynamo effect (based on the nonlinear time scale) and 
the return to equipartition due to the propagation of Alfv\'en waves: this theory leads to the correct relation between the slope of 
total energy spectrum and the residual energy spectrum.
It remains to explain how in turn the cascade can be modified by the amplitude of the magnetic excess, when it is imposed.

\acknowledgments
We thank G. Belmont, Y. Dong for several fruitful discussions. 
A. Verdini acknowledges partial funding from the European UnionÕs Seventh Framework Programme for research, technological development and demonstration under grant agreement no 284515 (SHOCK). Website: project-shock.eu/home/, and from the Interuniversity Attraction Poles Programme initiated by the Belgian Science Policy Office (IAP P7/08 CHARM).

\bibliography{grappin}

\appendix
\section{Reconciling SF and spectral properties}
We examine here the apparent contradiction between SF and spectral scaling properties.
We deal with \textit{global} SF, that are related with 1D reduced spectra by eq.~\ref{SPSF}.

\begin{figure*}[th]
\begin{center}
\subfigure{\includegraphics [width=0.33\linewidth]{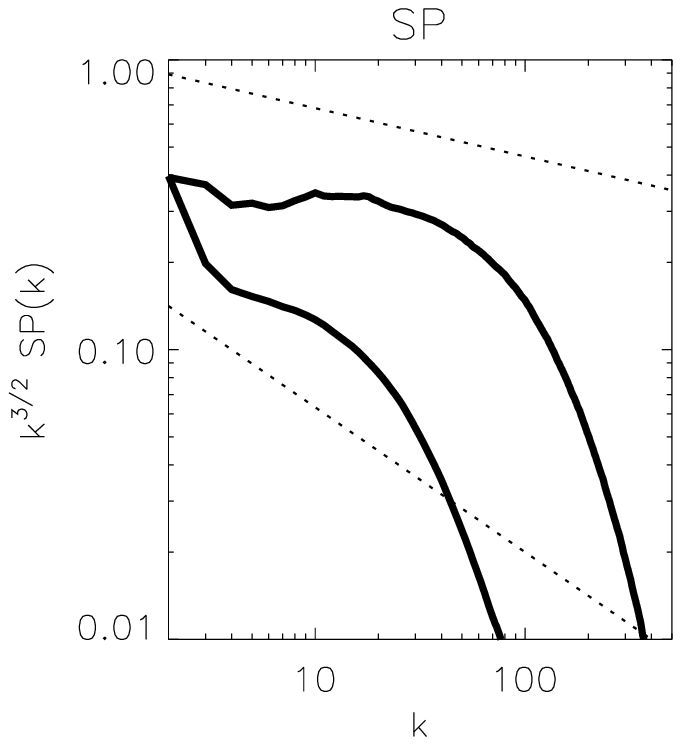}}
\subfigure{\includegraphics [width=0.33\linewidth]{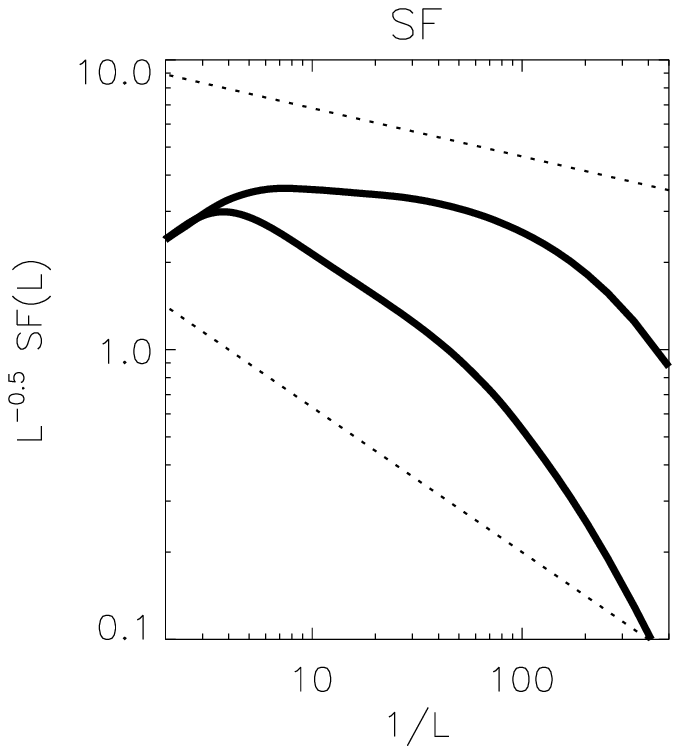}}
\\
\subfigure{\includegraphics [width=0.33\linewidth]{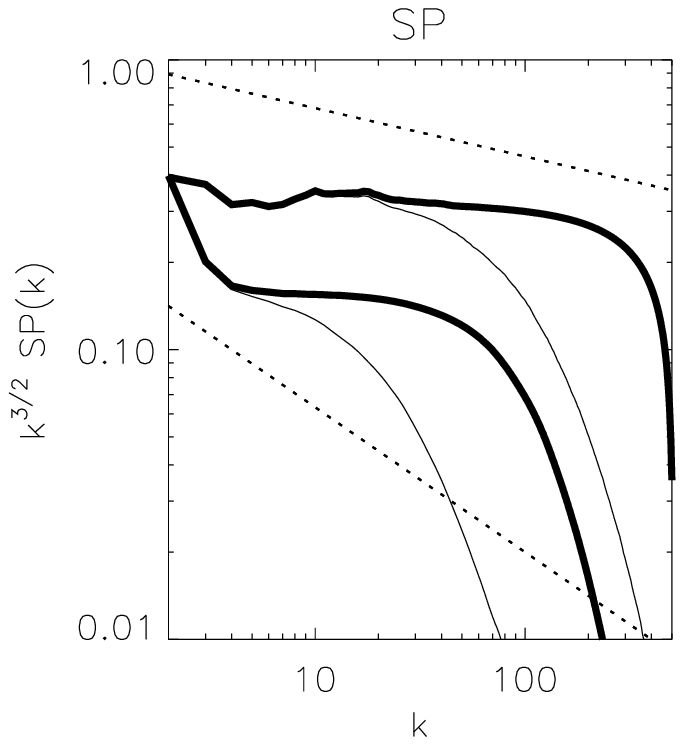}}
\subfigure{\includegraphics [width=0.33\linewidth]{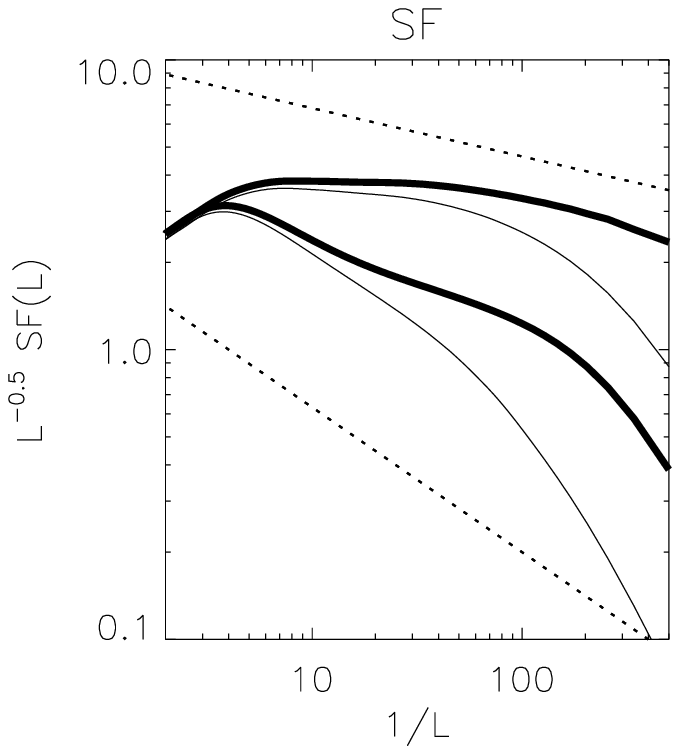}}
\subfigure{\includegraphics [width=0.33\linewidth]{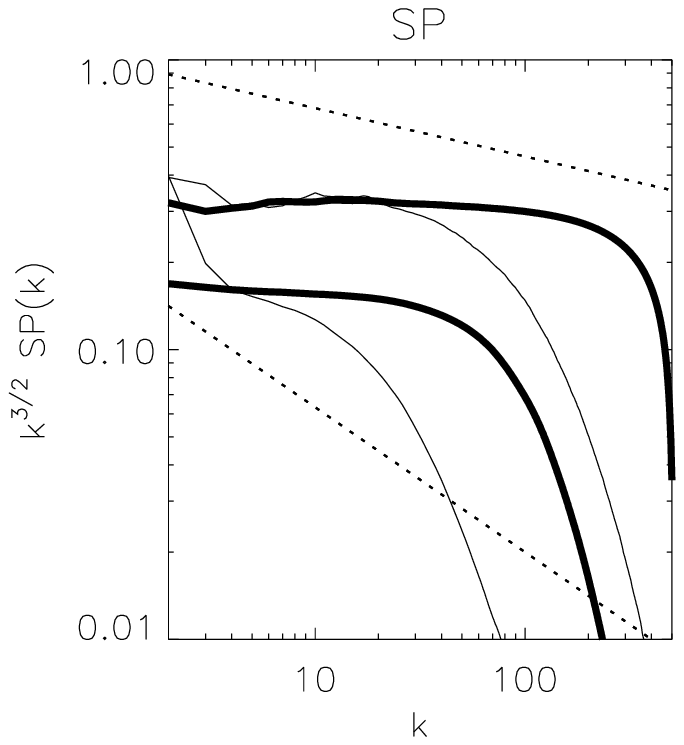}}
\subfigure{\includegraphics [width=0.33\linewidth]{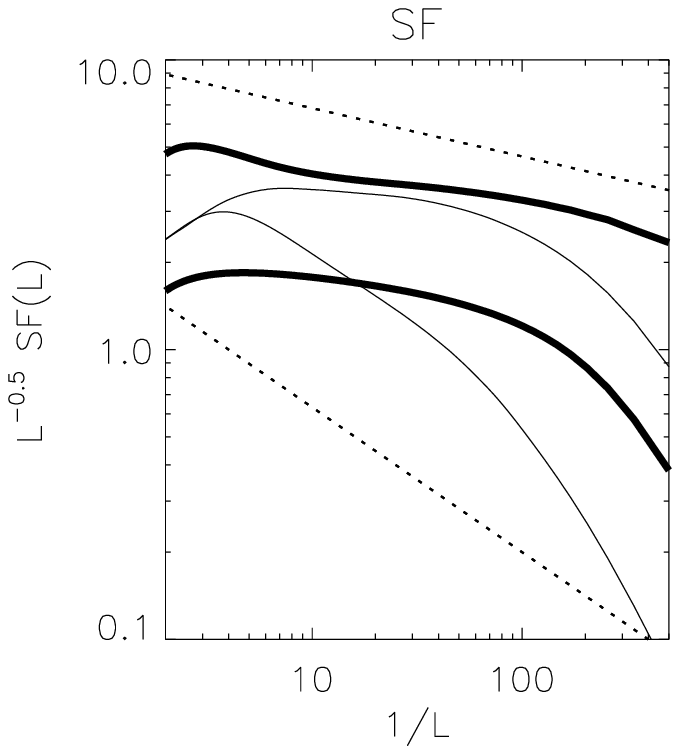}}
\caption{
Reduced 1D energy spectra SP (left, compensated by $k^{-3/2}$) and global second order structure functions SF (right, compensated by $L^{1/2}$), parallel (bottom curves) and perpendicular (upper curves),
(thick lines) built on three different versions of the gyrotropized 3D magnetic energy spectrum.
From top to bottom: (i) original spectrum, (ii) with small-scale scaling extrapolation, (iii) with large and small-scale scaling extrapolation.
Dotted lines for SF indicate $L^{2/3}$ and $L$ scalings; for SP they indicate $k^{-5/3}$ and $k^{-2}$ scalings.
Thin lines in middle and bottom raws indicate SP and SF built from the original 3D spectrum.}
\label{figSFa1}
\end{center}
\end{figure*}

A direct comparison of SP and SF is shown in fig.~\ref{figSFa1} (top raw). 
Perpendicular and parallel SP are obtained by taking a gyrotropic average of the 3D spectrum around the mean field direction. The SF are then deduced from the spectra by using the Fourier transform relation (eq.~\ref{SPSF}). (The reason for the gyrotropization will become clear below).
As seen immediately, the parallel spectral scaling is not measurable \citep{Muller:2005p705},
due to the limited scaling range at small angles with the mean field.
However, a parallel $k_\|^{-3/2}$ spectral scaling is easily made visible (middle raw) after performing a high Reynolds extrapolation of the gyrotropized 3D spectrum, by extending the $-3/2$ radial scaling range up to the maximum wavenumber for each angle $\theta$. 
Although the SP perpendicular and parallel scalings are identical now, the SF scalings remain
different in the two directions, although slightly closer one to the other than in the top raw.

The origin of the failure of the simple relation (eq.~\ref{dim}) between SF and SP scalings should thus lie in the large-scale structure, whose amplitude is isotropic in spectral space (as the forcing is), in contrast with the power-law range.
This hypothesis is confirmed by suppressing the forced range structure in the 3D spectrum, namely, by extending the $-3/2$ scaling to the \textit{whole} spectral range. 
As a result, we recover the expected correspondence between SF and SP scalings, as well as isotropic scalings, as shown in the bottom raw of fig.~\ref{figSFa1}.

\end{document}